\documentclass[%
 reprint, amsmath,amssymb, aps,]{revtex4-1}
\usepackage{graphicx}
\usepackage{dcolumn}
\usepackage{bm}
\usepackage{makeidx}
\usepackage{color}
\makeindex
\makeatletter

\newcommand{\Rmnum}[1]{\expandafter\@slowromancap\romannumeral #1@}
\makeatother
\begin{document}

\title{Multi-bit quantum digital signature based on quantum temporal ghost imaging}

\author{Xin Yao$^1$$^,$$^2$, Xu Liu$^1$$^,$$^2$, Rong Xue$^1$$^,$$^2$, Heqing Wang$^3$, Hao Li$^3$, Zhen Wang$^3$, Lixing You$^3$, Yidong Huang$^1$$^,$$^2$ and Wei Zhang$^{1,2}$}
\email{zwei@tsinghua.edu.cn}
\affiliation{$^1$Beijing National Research Center for Information Science and Technology (BNRist),\\ Beijing Innovation Center for Future Chips, Electronic Engineering Department, Tsinghua University, Beijing 100084, China\\
$^2$Beijing Academy of Quantum Information Sciences, Beijing 100193, China\\
$^3$State Key Laboratory of Functional Materials for Informatics, Shanghai Institute of Microsystem and Information Technology, Chinese Academy of Sciences, Shanghai 200050, China
}%

\begin{abstract}
Digital signature scheme is commonly employed in modern electronic commerce and quantum digital signature (QDS) offers the information-theoretical security by the laws of quantum mechanics against attackers with unreasonable computation resources. The focus of previous QDS was on the signature of 1-bit message. In contrast, based on quantum ghost imaging with security test, we propose a scheme of QDS in which a multi-bit message can be signed at a time. Our protocol is no more demanding than the traditional cases for the requirements of quantum resources and classical communications. The multi-bit protocol could simplify the procedure of QDS for long messages in practical applications. 
\end{abstract}

\maketitle

\section{Introduction}
Digital signatures aim to guarantee the authenticity and transferability of signed messages in modern digital communications. The classical protocols of digital signatures commonly employ the public-key encryption and provide the computational-assumption security for legitimate users. For example, the security of the famous Rivest-Shamir-Adleman protocol [{\color{blue}1}] is based on the reasonable capabilities in the computation of factorizing large integers, and such security is susceptible to algorithmic breakthroughs, large-scale computational resources and the emerging technologies of quantum computation [{\color{blue}2}-{\color{blue}4}]. In contrast, quantum digital signature (QDS), which was firstly proposed in 2001 [{\color{blue}5}], is robust against attackers with unrestricted capabilities in the computation, since QDS offers the information-theoretical security by fundamental principles of quantum mechanics. The early versions of QDS [{\color{blue}5},{\color{blue}6}] required the quantum memory for the interval between the distribution stage of quantum signature and the messaging stage, which is unfeasible in practical applications because of the immature technologies of quantum memory [{\color{blue}7},{\color{blue}8}]. Later, this demanding requirement was removed by using unambiguous state elimination for quantum states and only storing classical outcomes for the messaging stage [{\color{blue}9},{\color{blue}10}]. Wallden \emph{et al.} furthermore presented QDS protocols [{\color{blue}11}] requiring only the same technical components as quantum key distribution (QKD), which has been greatly developed over the past two decades [{\color{blue}12},{\color{blue}13}]. Such breakthroughs motivated several notable advances in experimental demonstrations of QDS [{\color{blue}14}], where the transmission distance has been remarkably extended by utilizing phase-encoded states [{\color{blue}15}-{\color{blue}17}] and polarization-encoded states [{\color{blue}18}-{\color{blue}20}] in recent years. Additionally, several proposals have been implemented for improving the security of QDS, such as measurement-device-independent (MDI) QDS [{\color{blue}21},{\color{blue}22}], which is immune to detector side-channel attacks, and passive decoy-state QDS for circumventing the leakage of the signal and decoy information to attackers [{\color{blue}23}]. However, those QDS protocols were dealing with the case of one-bit signature and the iterations of the procedure would be considerable for longer messages, limiting the feasibility in practical applications.
\par In this work, we propose and experimentally demonstrate a scheme of QDS in which the multi-bit message can be signed at a time. The multi-bit QDS protocol could markedly simplify the signature procedure for long messages in comparison with previous one-bit QDS. This work is inspired by ghost imaging in time domain [{\color{blue}24}-{\color{blue}27}] developed over the last several years. Ghost imaging is an intriguing technique of indirect imaging by the correlation of two beams. In a typical scheme of ghost imaging, one beam (so-called “test beam”) passing through the object is collected by the bucket detector without spatial resolution, while the multi-pixel detector which can image the object is placed in another spatially separated beam (“reference beam”). The first ghost imaging experiment was realized by the spatial correlation of photon pairs generated by spontaneous parametric down conversion (SPDC) in nonlinear crystals [{\color{blue}28}]. Thereafter, ghost imaging has been rapidly developed “from quantum to classical to computational” [{\color{blue}29}-{\color{blue}31}] with schemes of thermal-source ghost imaging [{\color{blue}29}], computational ghost imaging [{\color{blue}30}] and compressive ghost imaging [{\color{blue}31}]. Recently, Ryczkowski \emph{et al.} demonstrated a thermal ghost imaging scheme in time domain with an all-fiber setup [{\color{blue}24}]. Meanwhile, Zhang's group also realized quantum temporal ghost imaging and quantum secure ghost imaging over optical fibers of 50 km by utilizing time-frequency entanglement of photon pairs [{\color{blue}26},{\color{blue}27}], since this entanglement can be well maintained during the distribution of photon pairs over optical fibers. Quantum ghost imaging process can be treated as the transfer of multi-bit information between legal parties and suitable security test over the quantum channel can limit the attacker's knowledge on the imaging information, which can be applied in scenarios of quantum communications [{\color{blue}27}]. 
\par In this paper, we will firstly introduce the principle of quantum temporal ghost imaging based on time correlation of photon pairs, and then present the protocol and experimental demonstration of multi-bit QDS, in which the multi-bit message is signed by the way of ghost imaging. The multi-bit scheme would promote QDS toward practical applications by reducing the iteration of the signature procedure.

\section{Quantum temporal ghost imaging}
Fig.\,{\color{blue}1(a)} illustrates the typical scheme of quantum temporal ghost imaging. In this scheme, Energy-time entangled photon pairs are generated by the quantum light source placed at Alice. She keeps signal photons and sends idler photons to Bob over the quantum channel. At Alice’s side, signal photons pass through an intensity modulator followed by a single photon detector (SPD). The temporal object is the repeated time-varying pattern carried by the modulator. Similar to the temporal ghost imaging scheme in Ref.  [{\color{blue}24}], Alice’s detector has low resolution in time domain and therefore, it cannot image the temporal object. The resolution of Alice’s detector is equal to the repeated pattern’s period $T$. However, Bob’s detector has high resolution and it can effectively image the temporal object placed at Alice’s side. The timing clocks of the two SPDs are synchronized and the measurement time is equally divided into many frames (Fig.\,{\color{blue}1(b)}). Here the frame size is equal to the period of the temporal pattern $T$. Hence, Alice only knows the frames in which her detector collects signal photons, and she has no knowledge of the precise timing information of single photon events due to the low resolution of her SPD. On the other hand, Bob’s high-resolution SPD can record the precise arrival times of idler photons. After the measurement of millions of single photon events at both sides, Alice sends the frame numbers, labelled as $A_{\rm frame}$, to Bob by the classical channel. Then, Bob sifts his photon records $B_{\rm record}$ by keeping the records of photons in the same frames as $A_{\rm frame}$ and discarding others. Since the signal and idler photons of a pair are generated simultaneously from the entanglement source, the sifting operation with  $A_{\rm frame}$ actually informs Bob how the temporal pattern by the intensity modulator selects Alice’s photons. After the sifting, Bob could retrieve the pattern by making the statistics of photons’ positions in the corresponding frames. From the view of quantum ghost imaging, $A_{\rm frame}$ are the outputs of the bucket detector without temporal resolution in the test beam, while Bob’s precise timing information $B_{\rm record}$ are recorded by the detector with high resolution in the reference beam. Neither $A_{\rm frame}$ nor $B_{\rm record}$ can singly retrieves the temporal pattern. On the other hand, from the view of communications, Alice transmits the pattern information to Bob by the method of quantum ghost imaging. Therefore, the physical implementation of the intensity modulator can be equivalently replaced by the single photon event selection according to the repeated temporal pattern in Alice’s data processing, as illustrated in Fig.\,{\color{blue}1(c)}. In detail, at Alice’s side, signal photons are directly detected by a high-resolution detector and photons’ arrival times are precisely recorded. After the measurement, Alice selects the photon records according to a specific repeated pattern, which is the temporal object at Alice’s side. Then, Alice still sends the corresponding frame numbers $A_{\rm frame}$ to Bob through the classical channel. Obviously, for Bob and any other party, there is no difference between the typical scheme (Fig.\,{\color{blue}1(a)}) and the modified scheme (Fig.\,{\color{blue}1(c)}). In this paper, the modified scheme is adopted in the QDS protocol for its easier implementation comparing with the typical scheme with an intensity modulator. As mentioned above, quantum ghost imaging is based on the correlation measurement of two separate beam, hence, this technique actually realizes the transmission of the object information from Alice (test beam) towards Bob (reference beam). Furthermore, quantum temporal ghost imaging is based on photon pairs with energy-time entanglement. A part of entangled photon pairs could be utilized to monitor the security of the quantum channel in the imaging process [{\color{blue}27}]. Hence, quantum temporal ghost imaging can be applied as a new approach in quantum secure communications.

\begin{figure}
\includegraphics[scale=0.16]{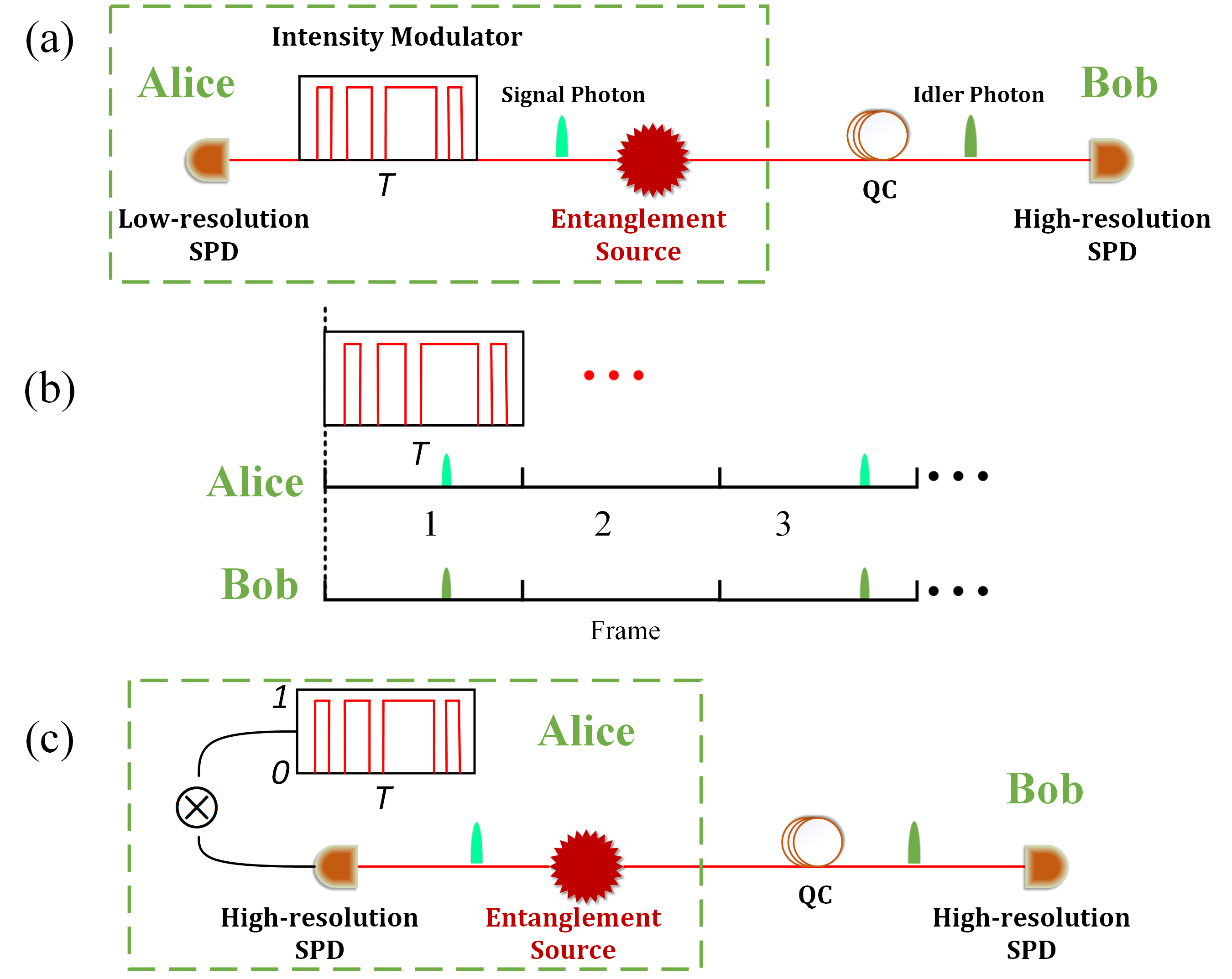}
\caption{(a) The sketch of quantum temporal ghost imaging; (b) Alice and Bob record the photons' arrival times and the corresponding frame numbers. SPD, single photon detector; QC, quantum channel.}
\label{fig:1}
\end{figure}
\par For reducing the background noise of temporal ghost imaging, we adopt the approach of large-alphabet QKD encoding [{\color{blue}32}] in which the photons' arrival times are encoded by three layers. A frame contains several time slots and a slot consists of a few bins. For the photon record in Fig.\,{\color{blue}2}, its frame number is 1, slot number is 1 and bin number is 4. In ghost imaging, the period of the binary pattern is as large as the frame size and the bit size of the pattern is the same as the slot size. The pattern displayed in Fig.\,{\color{blue}2} is “1001” and the four-bit message can be transferred from Alice to Bob by ghost imaging. To be specific, Alice and Bob firstly make the frame and bin sifting to reduce the effect of noise photons and detector jitter. The two parties publicly announce the photons' frame numbers and bin numbers, only keeping the records of those photons with same frame and bin numbers and discarding other records. Then, Alice utilizes the temporal pattern to select her single photon records and sends the corresponding frame numbers to Bob. The multi-bit message can be retrieved at Bob's side by the correlation of Alice's frame numbers and Bob's precise records.
\par In the three-layer encoding mechanism, Alice and Bob don't announce the slot numbers in the classical communication for ghost imaging, and the third party cannot eavesdrop the multi-bit message over the classical channel since the slot layer conveys the information of the temporal pattern. Furthermore, the security of quantum channel (Fig.\,{\color{blue}1(b)}) can be checked by the measurement of time-energy entanglement quality [{\color{blue}32},{\color{blue}33}] or the protocol of nonlocal dispersion cancellation [{\color{blue}34},{\color{blue}35}]. The upper bound of the eavesdropping fraction indicates the leakage of Bob's slot information and determines the security level of QDS protocol based on ghost imaging, which will be discussed in the following sections.
\begin{figure}
\includegraphics[scale=0.17]{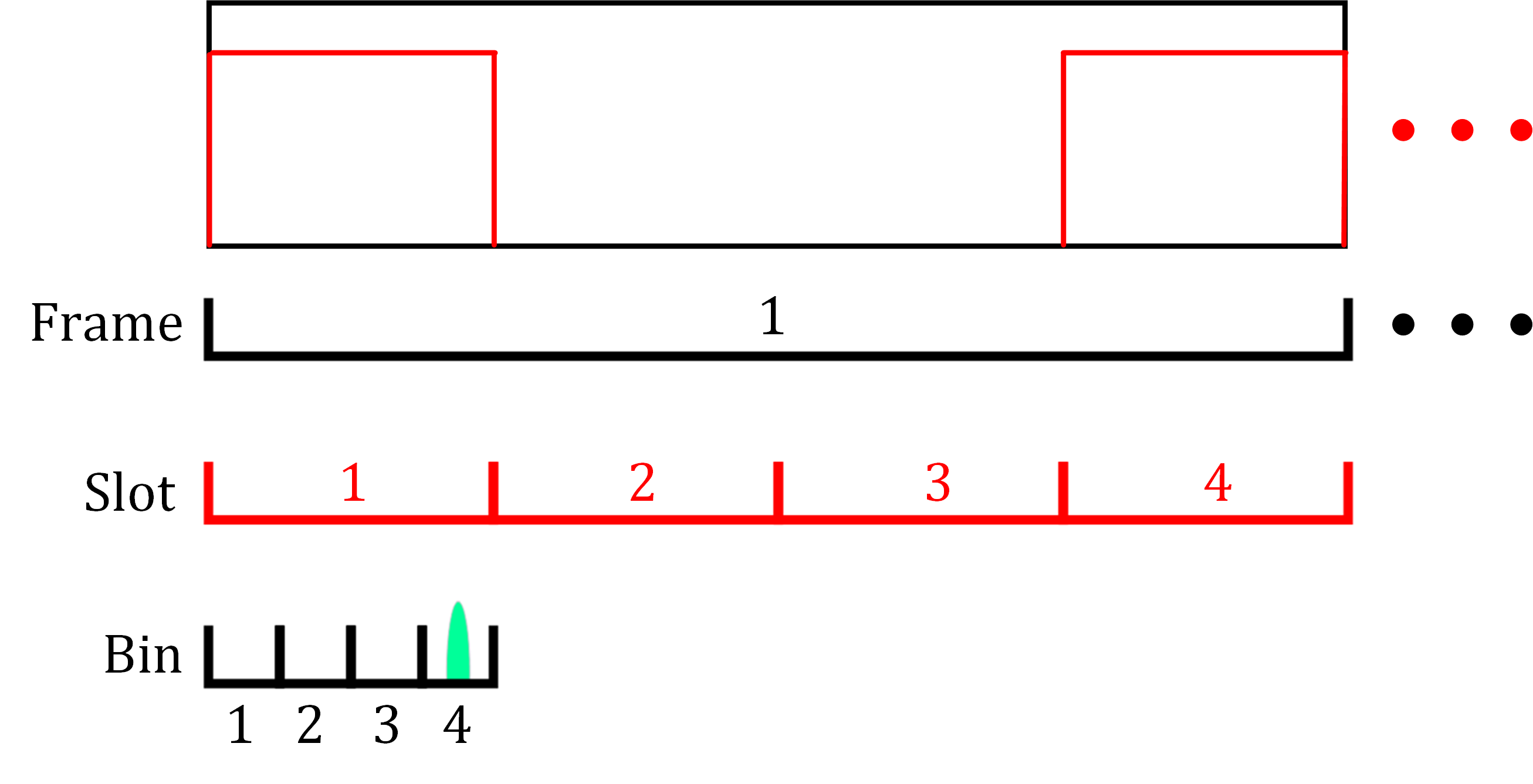}
\caption{Three-layer encoding in ghost imaging}
\label{fig:2}
\end{figure}

\section{protocol of multi-bit quantum digital signature}
There are generally three parties in the model of QDS wherein Alice signed the message and Bob (Charlie) are recipients. Fig.\,{\color{blue}3} illustrates the basic scheme of multi-bit QDS. Alice holds the entanglement source and keeps the signal photons of entangled pairs. Idler photons pass through a beam splitter and are distributed to Bob and Charlie over quantum channels. There are pairwise authenticated classical channels between Alice and Bob (Charlie), which can be realized with short preshared keys [{\color{blue}36}]. Additionally, there is a secure classical channel between Bob and Charlie, which can be guaranteed by the state-of-the-art technology of QKD. In QDS, Alice sends the message with her signature to Bob and Bob forwards the message to Charlie if he accepts it. QDS should be immune to Alice's repudiation and Bob's forging. A successful repudiation by Alice means Bob accepts the message but Charlie rejects it.

\par Similar to previous QDS, the multi-bit QDS has two stages: distribution stage and messaging stage. In the former stage, the quantum states are distributed to the parties and measured by them, while the latter stage corresponds to the transmission of the signed message by the classical communication.  
\par Basically, there are five steps in the distribution stage of the proposed protocol.
\par (1) Alice, Bob and Charlie implement the clock synchronization, and announce the sizes of time frame, slot and bin.
\par (2) The three parties measure the single photons, recording the arrival times of photons and the corresponding frame, slot and bin numbers.
\par (3) Alice and Bob (Charlie) publish part of records to estimate the error rate of slot encoding and the upper bound of the eavesdropping fraction $\chi_{\rm AB}$ ($\chi_{\rm AC}$) over the quantum channel $X$ ($Y$). A successful forging of the signed message is closely related to the considerable eavesdropping on photons toward the recipients of the signature, i.e., Bob and Charlie. Suitable security check can be utilized to monitor the quantum channel [{\color{blue}32}-{\color{blue}35}] and the success probability of the malicious forging exponentially increases as the eavesdropping fraction increases [{\color{blue}15},{\color{blue}16}]. 
\par (4) Alice and Bob announce the frame and bin numbers of the remaining photon records. They keep the records of photons in the same frames and bins, discarding other records. Then, Alice's records are labelled as $X^A$ and Bob's records as $X^B$ since their records correspond to the quantum channel $X$. The records here only contain the frame numbers and slot numbers of photons. Similarly, Alice and Charlie also make the frame and bin sifting. After the sifting, Alice's records are labelled as $Y^A$ and Charlie's records as $Y^C$ by quantum channel $Y$.
\par (5) Bob and Charlie secretly and randomly exchange half of photon records with each other to make the records symmetric from Alice's view. The secret exchange can be guaranteed by the classical secure channel between Bob and Charlie based on QKD technology. Similar to previous QDS, this step is to prevent Alice's repudiation. 
After the distribution stage, Bob has the records $S^B=(X^B_{\rm keep},Y^C_{\rm forward})$, wherein $X^B_{\rm keep}$ are the half of records Bob keeps in the secret exchange and $Y^C_{\rm forward}$ are the records forwarded from Charlie. Also, Charlie's records are denoted as $S^C=(X^B_{\rm forward},Y^C_{\rm keep})$. Alice's records are $S^A=(X^A,Y^A)$.
\begin{figure}
\includegraphics[scale=0.35]{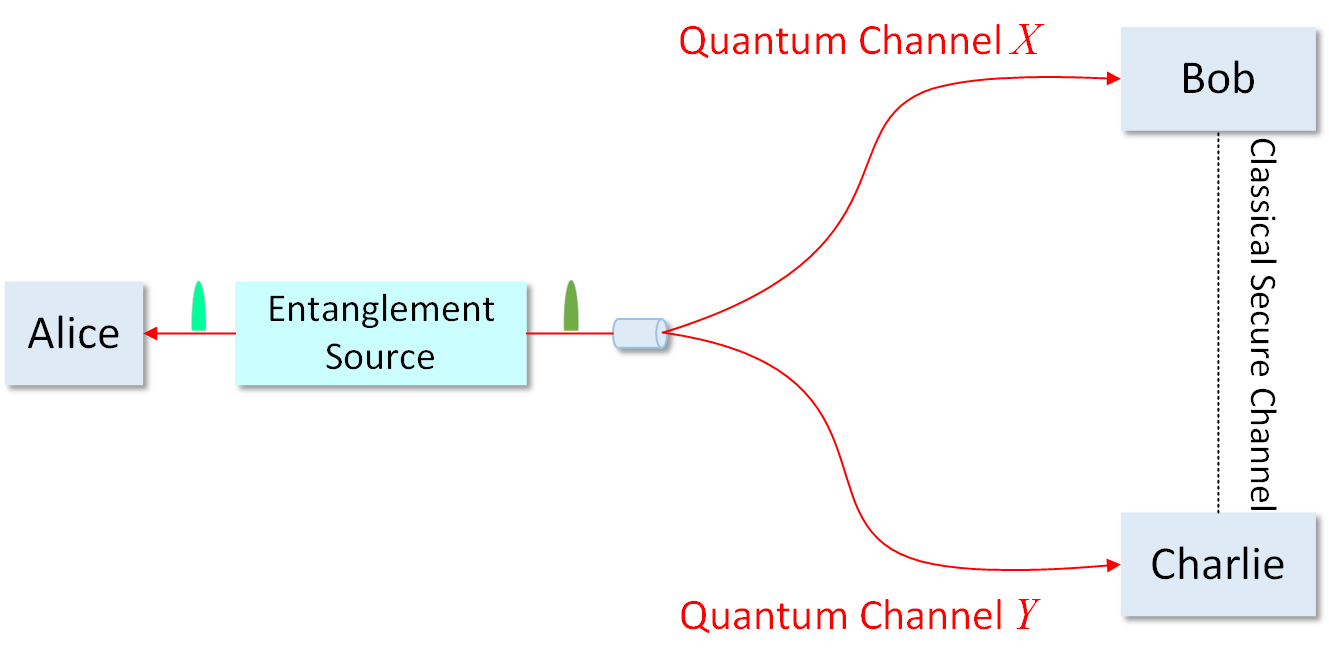}
\caption{The setting of multi-bit QDS. Alice holds the entanglement source and keeps signal photons, while idler photons are randomly directed to Bob and Charlie by a beam splitter.}
\label{fig:3}
\end{figure}
\par The messaging stage is the classical communication, which could occur much later. There are three steps in this stage.
\par (6) Alice randomly keeps half the records $S^A=(X^A,Y^A)$ and discards others. Thereafter, Alice chooses the records with specific slot numbers according to the message to be signed, and then sends the frame numbers of those records she chooses to Bob through the classical channel. Here the frame numbers are the signature elements, denoted as $Sig_{\rm frame}$. If a frame consists of M slots, M-bit signed message has been transmitted.
\par (7) When Bob receives the frame numbers $Sig_{\rm frame}$, he can retrieve the M-bit signed message by the temporal ghost imaging, as discussed in Section \Rmnum{1}. Bob has two record blocks $X^B_{\rm keep}$ and $Y^C_{\rm forward}$, and hence he can perform ghost imaging twice. The first ghost imaging is realized by $X^B_{\rm keep}$ and $Sig_{\rm frame}$, and the second one is by $Y^C_{\rm forward}$ and $Sig_{\rm frame}$. The noise of ghost imaging is due to system error rate, which includes setup imperfection and the eavesdropping perturbation over the quantum channel, and the forging of signature numbers $Sig_{\rm frame}$. The low leakage of slot information would lead to a high noise of ghost imaging in case of forging the frame numbers and vice versa. If both of the noise factors in the two ghost images, defined as the inverse of signal-to-noise ratio, are lower than the acceptance threshold $Th^B_{\rm accept}$, Bob will accept the message and forwards it to Charlie through the classical channel. The noise factor here is similar to the concept of \emph{mismatch} in previous QDS [{\color{blue}15}-{\color{blue}23}] where Bob checks Alice's raw keys to decide whether he accepts the message or not. 
\par (8) Similarly, Charlie uses his records $X^B_{\rm forward}$ and $Y^C_{\rm keep}$ to perform ghost imaging twice with the forwarded signature elements $Sig_{\rm frame}$. Also, if both of the noise factors in the two images are lower than the verification threshold $Th^C_{\rm verify}$, Charlie will accept the message. It should be noted that Charlie's threshold is higher than Bob's, i.e., $Th^B_{\rm accept}<Th^C_{\rm verify}$, to prevent Alice's repudiation.
\par It's worth noting that in our protocol two special cases should be avoided: all bits of the signed message are 1 or 0, since Bob could easily forge any message to be “1111…” or “0000…” according to the frame sifting of Alice and Charlie in the classical channel.
\section{Proof-of-principle demonstration}
\begin{figure*}
\includegraphics[scale=0.22]{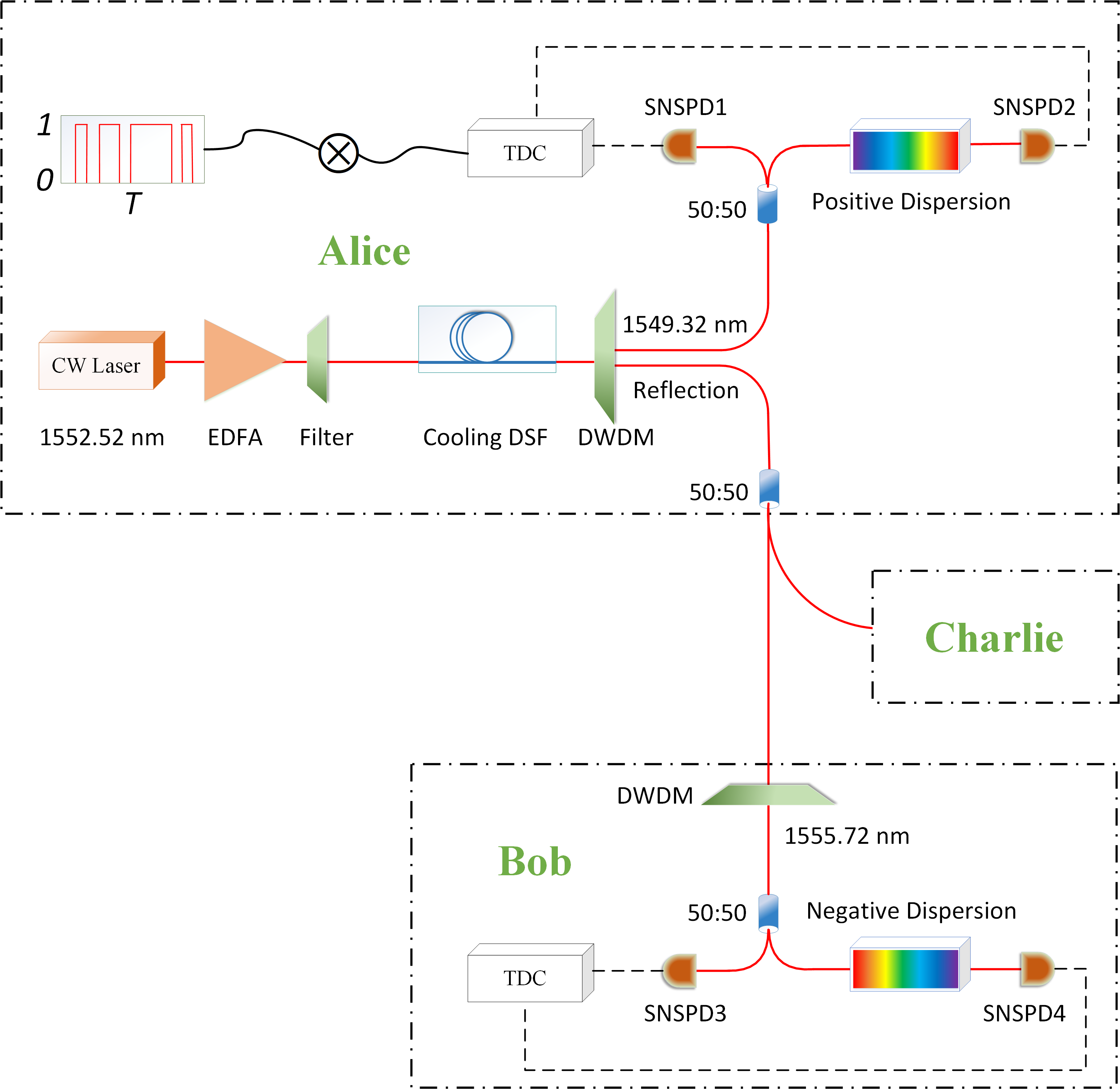}
\caption{Experimetal setup. The time-energy entangled photon pairs are generated in the cooling dispersion shifted fiber (DSF). Alice holds the quantum source and keeps the signal photons of pairs, while idler photons are collected by Bob and Charlie. Charlie's setup is identical to Bob's. Positive and negative dispersion modules are used for the measurement of nonlocal dispersion cancellation to estimate the eavesdropping fraction of the quantum channel between Alice and Bob (Charlie). EDFA: erbium doped fiber amplifier; DWDM, dense wavelength division multiplexer; SNSPD, superconducting nanowire single photon detector; TDC, time-to-digital converter. }
\label{fig:4}
\end{figure*}
\par The experimental setup of multi-bit QDS is shown as Fig.\,{\color{blue}4}. A continuous-wave laser with center- wavelength of 1552.52 nm pumps a piece of dispersion shifted fiber (DSF) to generate time-energy entangled photon pairs by the spontaneous four-wave mixing process. The length of DSF is about 200 meters and it is cooled at about 40 K in the Gifford-Mcmahon cryocooler for suppressing Raman noise photons. The generated photon pairs are filtered by a dense wavelength division multiplexer (DWDM) with center-wavelength of 1549.32 nm and bandwidth of 50 GHz. The signal photons are collected by Alice, while the reflected photons pass through the 50:50 coupler and are distributed to Bob and Charlie. At Bob side, he uses another DWDM with center-wavelength of 1555.72 nm and bandwidth of 50 GHz to collect the idler photons. Charlie's setup is identical to Bob's and therefore is not shown in detail. The fiber distance between Alice and Bob (Charlie) is a few meters. For the three parties, they all directly detect half the daughter photons to estimate the channel error rate and perform the digital signature based on ghost imaging, while the other half of photons at each side pass through the positive (negative) dispersion module to estimate the security of quantum channel by the nonlocal dispersion cancellation [{\color{blue}34},{\color{blue}35},{\color{blue}37}]. The positive (negative) dispersion module (DCMCB, Proximion Corp.) is based on the fiber Bragg grating with group velocity dispersion of 1981 (-1980) ps/nm at 1545nm. The detection efficiencies of superconducting nanowire single photon detectors (SNSPD) are $\sim$50\% with dark count rates of $\sim$100 Hz, timing jitters of $\sim$80 ps and maximum count rates of $\sim$2 MHz [{\color{blue}38}]. The coincidence measurements of the single photon events are realized by a time-to-digital converter (TDC) modules (Hydra Harp 400, Pico Quant). The full width at half maximum (FWHM) of the coincidence peak between SNSPD1 and SNSPD3 is 128 ps, while coincidence between SNSPD2 and SNSPD3 manifests the large dispersion effect with the FWHM of 896 ps. The peak of nonlocal dispersion cancellation is 160 ps by the coincidence measurement of SNSPD2 and SNSPD4. The eavesdropping of the collective-attack level on photons' timing information would be indicated by the nonlocal dispersion cancellation [{\color{blue}34},{\color{blue}35}].

\par In the experiment, the photons' timing information are encoded by the three-layer mechanism as shown in Fig.\,{\color{blue}3}. Considering the filtering bandwidth of photon pairs is 50 GHz, the size of bin is set as 20 ps, similar to Ref. [{\color{blue}34}]. In detail, a time slot contains 15 bins and a frame consists of 10 slots. Therefore, the size of the frame is 3 ns. Taking into account of the single count rate of Alice's detector SNSPD1 as 1.5 MHz, there is basically no more than one photon in each frame and multi-photon records are discarded. In the step (5) of distribution stage, Bob and Charlie secretly and randomly exchange half of the records, i.e., frame and slot numbers of photons, with each other through the classical secure channel. After this stage, Alice holds the photon records $S^A=(X^A,Y^A)$, where $X^A$ ($Y^A$) corresponds to quantum channel $X$ ($Y$) between Alice and Bob (Charlie). Bob has the records $S^B=(X^B_{\rm keep},Y^C_{\rm forward})$, where $X^B_{\rm keep}$ are the records he keeps in the exchange and $Y^C_{\rm forward}$ are forwarded from Charlie. Similarly, Charlie's records are $S^C=(X^B_{\rm forward},Y^C_{\rm keep})$ where the former are from Bob and the latter are kept by Charlie himself. 
\begin{figure}
\includegraphics[scale=0.21]{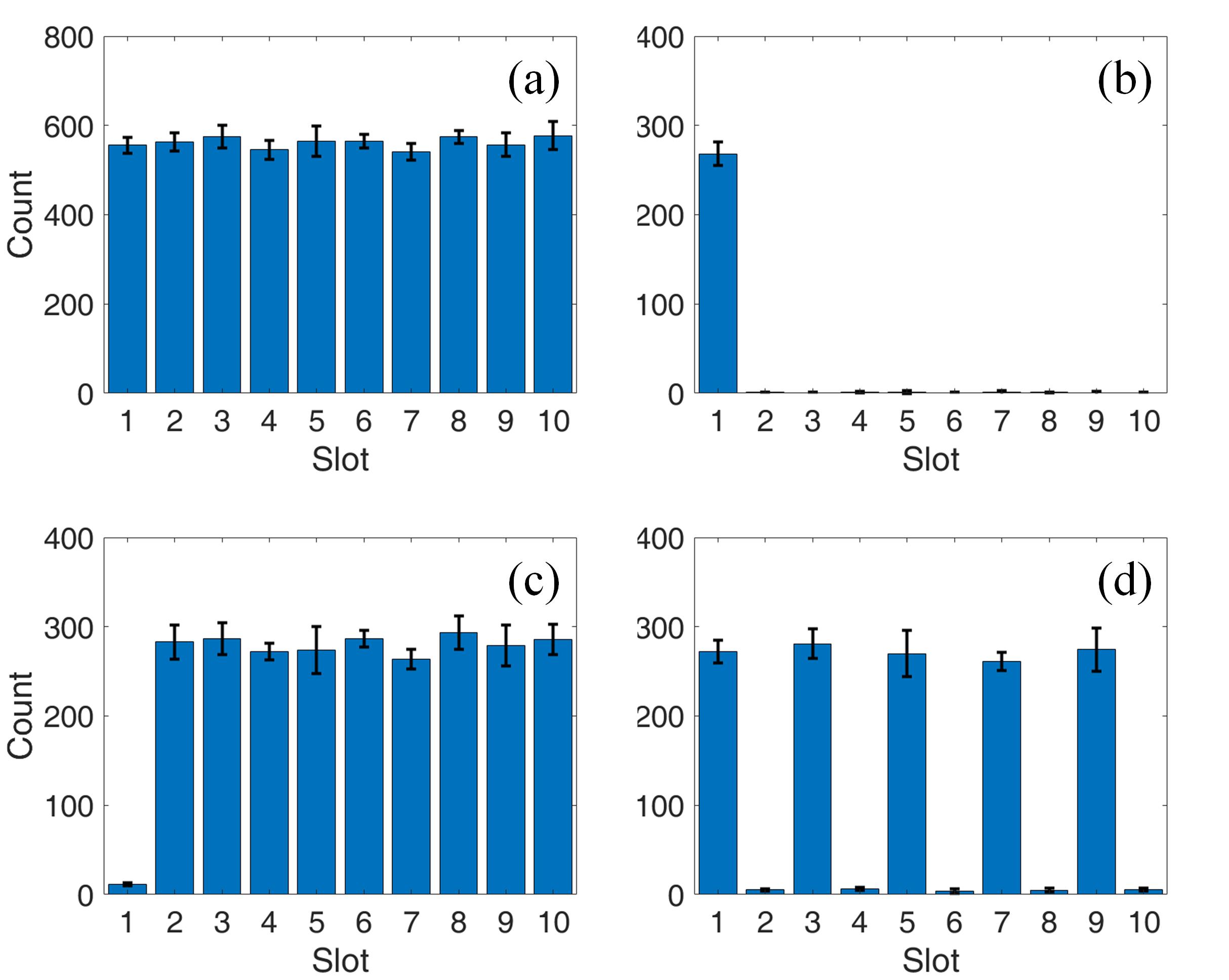}
\caption{(a) The distribution of Bob's records $X^B_{\rm keep}$ in each slot after the distribution stage; Bob's ghost images after Alice signs (b) “1000000000”, (c) “0111111111” and (d) “1010101010” in the messaging stage.}
\label{fig:5}
\end{figure}

\par Next, we take the data block $X^B_{\rm keep}$ as an example to explain the messaging stage of 10-bit QDS by ghost imaging. 10-bit message can be signed at a time since a frame contains 10 slots in our setting. Fig.\,{\color{blue}5(a)} presents the distribution of Bob's records $X^B_{\rm keep}$ in each slot with the single-photon measurement time of 2 s.The average count of $\langle X^B_{\rm keep}\rangle$ is 561.93$\pm$7.26 and the error rate of slot number between Alice and Bob is 3.78\%$\pm$0.16\%. For signing the message “1000000000”, Alice firstly randomly chooses half the records from  $S^A=(X^A,Y^A)$ and then selects the records of which the slot numbers are all 1. Thereafter, she sends the frame numbers of those records, i.e., the signature elements $Sig_{\rm frame}$, to Bob. Finally, Bob retrieves the message by ghost imaging of $X^B_{\rm keep}$ and $Sig_{\rm frame}$ (Fig.\,{\color{blue}5(b)}). Since Alice randomly discards half of photon records at the beginning of the messaging stage, the count in the first slot is 268.33$\pm$13.31, which is approximately the half of $\langle X^B_{\rm keep}\rangle$. Therefore, in the messaging stage, if $X^B_{\rm keep}(i)$ in the \emph{i}th slot is not less than $\langle X^B_{\rm keep}\rangle/2-\sigma$, where $\sigma=\sqrt{\langle X^B_{\rm keep}\rangle/2}$ is for the account of Poisson fluctuation, the bit in this slot can be judged as “1”. If not, the bit is “0”. In Fig.\,{\color{blue}5(b)}, the counts in other slots are due to the system error rate between Alice's records with slot numbers of 1 and Bob's records with slot numbers of 2 to 9. Fig.\,{\color{blue}5(c)} is Bob's image corresponding to the message “0111111111”. The count in the first slot is 10 due to the system error rate between Alice's records with slot numbers of 2 to 9 and Bob's records with slot numbers of 1. Here we define the noise factors in the slots of bit “0” as $f(X^B_{\rm keep}(i))\equiv X^B_{\rm keep}(i)/(\langle X^B_{\rm keep}\rangle/2)$, where \emph{i} is slot number. For the first slot of Fig.\,{\color{blue}5(c)}, the noise factor $f(X^B_{\rm keep}(1))=4.09\%$$\pm$0.59\%, close to the system error rate of 3.78\%$\pm$0.16\%. In real applications, the increasing noise factor is attributed to the perturbation by the eavesdropping in the distribution stage and the malicious forging in the messaging stage. Actually, the noise factor can be treated as the \emph{mismatch} of signature elements in the traditional QDS [{\color{blue}15}-{\color{blue}23}]. On the other hand, Bob has an acceptance threshold $Th^B_{\rm accept}$. After receiving the signature elements $Sig_{\rm frame}$, if all the noise factors in each slot of the two ghost images ($Sig_{\rm frame}$ and $X^B_{\rm keep}$, $Sig_{\rm frame}$ and $Y^C_{\rm forward}$) are less than the threshold $Th^B_{\rm accept}$, Bob will accept the message. The value of $Th^B_{\rm accept}$ should be slightly larger than the system error rate. Furthermore, Bob forwards $Sig_{\rm frame}$ to Charlie and Charlie will also accept the message if the noise factors are less than his threshold $Th^C_{\rm verify}$. By this step, a 10-bit QDS is realized. To prevent Alice's repudiation, $Th^B_{\rm accept}$ is smaller than $Th^C_{\rm verify}$, i.e., Charlie will certainly accept the message if Bob accepts it. Fig.\,{\color{blue}5(d)} is the case of the signed message “1010101010”.

Finally, the security level can be calculated in the 10-bit QDS. As the noise factor can be treated as the \emph{mismatch} in previous QDS schemes, the security-level equations in Ref.\,[{\color{blue}16}] can be adopted in the multi-bit model as:
\begin{equation}
Prob(Honest \ Abort)\le2{\rm Exp}[-(Th^B_{\rm accept}-e)^2L];
\tag{1a}
\end{equation}
\begin{equation}
Prob(Repudiation)\le2{\rm Exp}[-(\frac {Th^C_{\rm verify}-Th^B_{\rm accept}}{2})^2L];
\tag{1b}
\end{equation}
\begin{equation}
Prob(Forge)\le{\rm Exp}[-(P_e-Th^C_{\rm verify})^2L].
\tag{1c}
\end{equation}

In the experiment, the system error rate $e=3.78\%$ and the average count in ghost imaging is $L=\langle X^B_{\rm keep}\rangle/2+\langle Y^C_{\rm forward}\rangle/2$ with the count rate of 276 per second. $P_e=1-\chi_{\rm AC}$ is Bob's probability of incorrectly guessing the slot numbers of Charlie's records $Y^C_{\rm keep}$ when Bob wants to forge the message. Here we set $P_e=0.447$ for the eavesdropping at the collective-attack level [{\color{blue}35}]. In general, the security level $\varepsilon$, which means the failure possibility of the protocol, is the maximum value of the three possibilities described above. Fig.\,{\color{blue}6(a)} displays $\varepsilon$ versus the Bob's and Charlie's thresholds when $L=552$, corresponding to the measurement time of 2 s in the distribution stage. We can also obtain the dependence of the optimized value $\varepsilon$ on the count $L$ in the ghost imaging (Fig.\,{\color{blue}6(b)}). For the QDS with typical security level of $10^{-4}$, $L$ is taken as 939. Bob's threshold $Th^B_{\rm accept}$ is set as 0.1410 and Charlie's threshold $Th^C_{\rm verify}$ is 0.3474. Hence, we have the security level $\varepsilon=0.91\times10^{-4}$. Therefore, the three parties need to make the single photon measurement for 4 s to control the failure probability below $10^{-4}$ in the 10-bit QDS. 
\begin{figure}
\includegraphics[scale=0.1825]{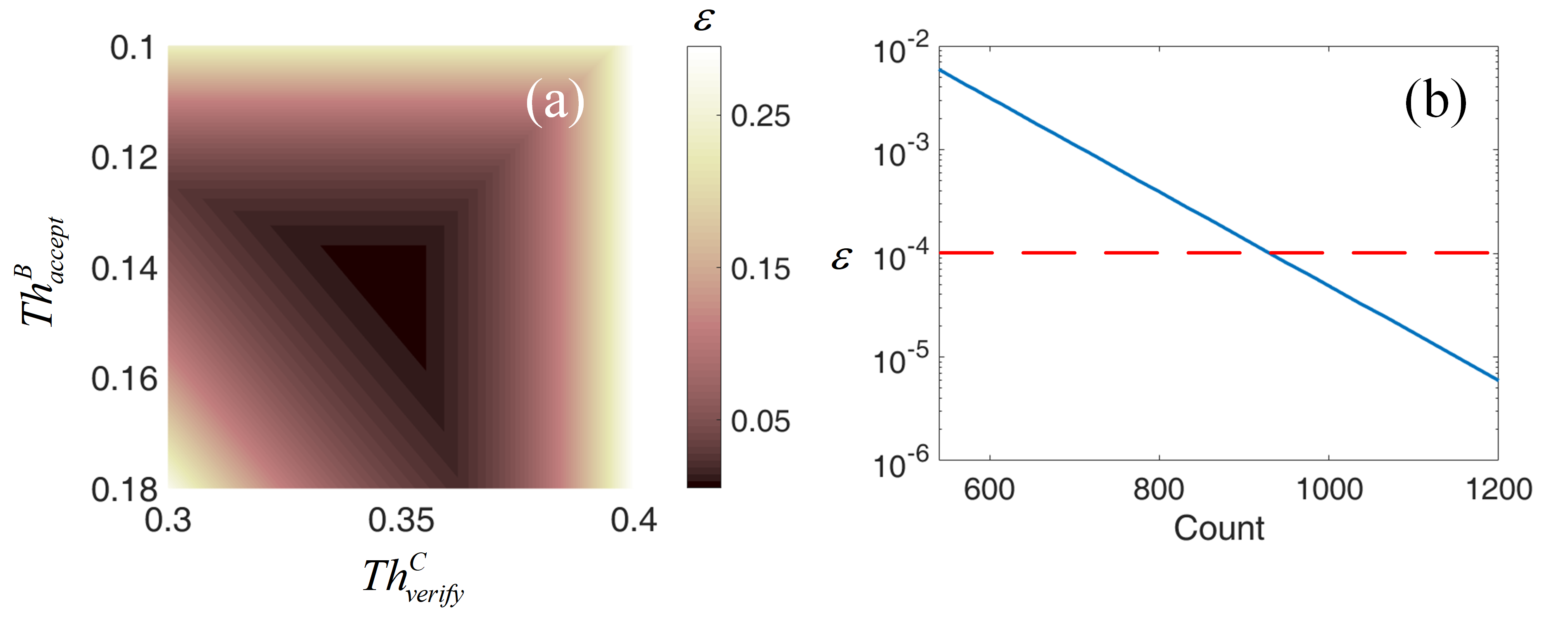}
\caption{(a) The security level $\varepsilon$ versus Bob's and Charlie's thresholds ( $Th^B_{\rm accept}$ and $Th^C_{\rm verify}$); (b) the optimized value of $\varepsilon$ versus the count $L$.}
\label{fig:6}
\end{figure}

\section{Discussion}
It is worth giving a comparison between the proposed mutli-bit QDS protocol and previous protocols when they are applied to sign long massages. Previous QDS protocols [{\color{blue}15}-{\color{blue}23}] focused on the single-bit signature. For a long message in practical applications, it should be signed bit by bit since these QDS protocols are based on the raw keys generated by QKD for each bit. Compared to the iteration procedure of the single-bit signature protocols when they are applied on signing long messages, the proposed multi-bit signature protocol has two characteristics:
\par (a) Eliminating the impact of selective attacks, in which one bit or a few specific bits in a multi-bit massage are attacked.
\par When the single-bit QDS protocol is applied on signing a long massage, the key strings of one bit or a few specific bits in the long massage may be attacked selectively in the distribution stage of QDS. Hence, the length of the key string for these bits should be large enough in order to guarantee its security. On the other hand, investigations have shown that the iteration procedure of single-bit QDS protocols requires that the key strings generated in the distribution stage for different bits in the massage should be consumed by sequence to avoid Bob’s malicious forging, and the length of the key string for each bit should be equal [{\color{blue}39}-{\color{blue}41}]. It can be seen that, to avoid the selective attack on a specific bit, the key strings for all the bits in a long massage should be long enough, although most of them might not be attacked. Hence, it reduces the efficiency of quantum sources, i.e. the sources for the generation of raw key strings. 
\par In the proposed multi-bit QDS protocol based on quantum ghost imaging, the retrieve of the multi-bit message is based on the data blocks from ghost-imaging detectors at both sides. Since the photon pairs are generated randomly in time domain, and arrival times of single photon event recorded at Bob and Charlie are used to retrieve all the bits of the long massage. Hence, the proposed multi-bit QDS protocol eliminates the impact of selective attacks, in which one bit or a few specific bits in a multi-bit massage are selectively attacked. 

\par (b) Simplifying the security-check procedure.
\par When the single-bit QDS protocol is applied on signing a long massage, the key strings for each bit should be carefully checked for security against eavesdropping. The participants should check them one by one, and could not use a global security-check parameter for the whole massage in the distribution stage to estimate the eavesdropping bound, since the semantic difference of the message could be modified by forging a few bits or only one single bit.
\par  In the proposed multi-bit QDS protocol, the security check in the distribution stage is global for the whole massage. It can be used to estimate the attacker’s eavesdropping, since photon pairs contribute to each bit with equal possibility and the protocol can avoid the selective attack intrinsically. Hence, the security-check procedure is obviously simplified by the proposed protocol in the case of long massages. 

\section{Conclusion}
We have presented a protocol of multi-bit QDS with a proof-of-principle demonstration in which the 10-bit message can be signed at a time. The message transmission is actually based on the quantum temporal ghost imaging, which is compatible with the security mechanism of nonlocal dispersion cancellation in the large alphabet-QKD scheme. In the distribution stage, the parties implement the measurements of the entangled photon pairs and store the classical measurement outcomes after the security check of pairwise quantum channels and necessary sifting on single photon events. In the messaging stage, the recipients, Bob and Charlie, retrieve the message by ghost imaging between their photon records and the forwarded signature elements from Alice. Bob (Charlie) accepts (verifies) the message by checking whether the noise factors in ghost images are below the acceptance (verification) thresholds. In real applications, the noise may due to the perturbation of the attacker's eavesdropping in the distribution stage and the forging of the forwarded signature elements in the messaging stage. The noise factor in our scheme is similar to the mismatch of raw-key sequence in traditional QDS and hence the security level of multi-bit protocol can be calculated by the approaches discussed in previous cases. In the experiment, 10-bit QDS is implemented with the distribution stage of 4 s and the security level of $10^{-4}$. The volume of 10 bit could be furthermore increased by suitably setting the parameters in three-layer encoding mechanism, simplifying the procedure of signing long messages in practical applications.

\section{Acknowledgments}
This work was supported by the National Key R\&D Program of China under Contract No. 2017YFA0303704 and No. 2017YFA0304000; the National Natural Science Foundation of China under Contract No. 61575102, No. 91750206, No. 61671438 and No. 61621064; Tsinghua National Laboratory for Information Science and Technology.

\end{document}